\documentclass[pdftex,twocolumn,epjc3]{svjour3}          
\pdfoutput=1
\RequirePackage[T1]{fontenc}

\smartqed  

\usepackage{graphics}
\usepackage{graphicx}
\usepackage{float}
\usepackage[colorlinks,citecolor=blue,urlcolor=blue,linkcolor=blue]{hyperref}
\usepackage[lofdepth,lotdepth]{subfig}
\usepackage[numbers,sort&compress]{natbib}
\usepackage{mciteplus}
\journalname{Eur. Phys. J. C}

\begin{document}

\title{Interacting Holographic Generalized Chaplygin Gas in compact Kaluza-Klein cosmology}


\author{S. Ghose\thanksref{e1,addr2,addr3}
        \and
        A. Saha\thanksref{e2,addr1} 
				\and
				B. C. Paul\thanksref{e3,addr3}
}

\thankstext{e1}{e-mail: souviknbu@rediffmail.com}
\thankstext{e2}{e-mail: arindamjal@gmail.com}
\thankstext{e3}{e-mail: bcpaul@iucaa.ernet.in}

\institute{Jalpaiguri Engineering College, Jalpaiguri, West Bengal, India \label{addr1}
          \and
          Physics Department, University of North Bangal, Rajarammohunpur,Siliguri, West Bengal, India 734013\label{addr2}
          \and
          \emph{Present Address:} Surendra Institute of Engineering and Management, Dagapur, Siliguri, West Bengal, India\label{addr3}
}

\date{Received: date / Accepted: date}

\maketitle

\begin{abstract}
We investigate  Holographic Dark Energy Correspondence of Interacting Generalized Chaplygin Gas  model in the framework of compact Kaluza-Klein cosmology. The evolution of the modified holographic dark energy and  the equation of state parameter is obtained here.Using the present observational value of density parameter a stable configuration is formed which accommodates Dark Energy. we note a connection between Dark Energy and Phantom field and have shown that Dark Energy might have evolved from a Phantom state in the past.

\end{abstract}

\section{Introduction}
Contemporary cosmological observations predict that we live in an accelerating universe \cite{riess,*perlm,*perlm2,*perlm3}. It remains to be understood, how to incorporate the observational predictions naturally in the standard frame work of cosmology. It is also predicted that $73 \%$ of the matter-energy content of the universe is due to something which has a negative pressure and commonly known as Dark Energy (DE). However, a tiny cosmological constant may be useful to address the observational data satisfactorily. Unfortunately while this solves the acceleration problem fairly well, it evokes other problems, e.g. the problem of fine tuning and cosmic coincidence problem etc. However,  the above accelerating universe problem may be addressed in two ways: (i) considering modified theory of gravity, i.e. modifying the gravity sector or (ii) considering some unusual form of matter/energy so as to yield an accelerating phase in late universe, i.e. modification of the matter sector. In the former case a number of theories of gravity, for example with $f(R)$ \cite{carroll}, $f(T)$ \cite{uddin} or $f(G)$ \cite{linderfg} are proposed. A large number of other matter fields e.g. quintessence \cite{samipaddy}, tachyon \cite{gibb} etc. are considered to modify the gravitational sector. 
Chaplygin gas (CG) was first introduced in 1904 in aerodynamics but recently it is being considered as one of the prospective candidate for dark energy. Although it contains a positive energy density it is referred as an exotic fluid due to its negative pressure. CG may be described by a complex scalar field originating from generalized Born-Infeld action. The equation of state for CG is given by:
\begin{equation}
\label{cgeos}
p=-\frac{A}{\rho},
\end{equation}
where $A$ is a positive constant. Later, a modified form of CG was considered \cite{billic, bentosen}. The equation of state for the modified CG is given by:
\begin{equation}
\label{gcgeos}
p=-\frac{A}{\rho^{\alpha}},
\end{equation}
where $0 \leq \alpha \leq 1$. This modified fluid is known as Generalized Chaplygin Gas (GCG). At high energy GCG behaves almost like a pressure less dust where as at low energy regime it behaves like a dark energy candidate its pressure being negative and almost constant. Thus GCG smoothly interpolates between a non-relativistic matter dominated phase in the early universe and a dark energy dominated phase in the late universe. This interesting property of GCG has motivated people to consider it as a candidate for unified dark matter and dark energy models on the other hand modification of the underlying theory of gravitation, however, can be thought of from a fundamentally different perspective. One of the interesting approaches is the theory of extra dimensions due to Kaluza \cite{kaluza} and Klein \cite{klein}. They used a $5D$ spacetime to unify gravity and electromagnetism. Kaluza-Klein (KK) model has been used in many literature for studying the models of cosmology as well as particle physics \cite{leekk, appelkk}. The cosmology of $5$-dimensional with pure geometry has a feature that properties of matter in $4$ dimension is induced by the $5$D geometry itself by virtue of what is known as Campbell's Theorem \cite{liucamp}. However, KK cosmology is also studied by including matter as well \cite{faraz, darabi}
Recently another very interesting approach  in quantum gravity, namely the holographic principle \cite{susskind}, is used to understand the DE problem. According to this principle:{\it all information inside any spatial region can be observed, instead of its volume, on its boundary surface}. It was shown by Cohen. {\it et. al.} \cite{cohen} that quantum zero point energy of any system of size $L$ (its Infra Red (IR) cutoff) can not exceed the mass of a black hole of the same size. So, if $\rho_v$ denotes the quantum zero point energy density and $M_P$ be the reduced Planck Mass $L^3 \rho_v \leq L M_P^2$.  A number of literature appeared \cite{cohen, li} where it was discussed how this Holographic principle could set up a proper ultraviolet cut off for the quantum zero point energy of the universe which in turn prevents one from grossly overestimating the value of $\Lambda$ from field theoretic  consideration. Very recently the idea of Holographic Dark Energy (HDE) has been incorporated to build an interacting dark energy model in KK cosmology \cite{sharifkk}. In this letter, using the above theory we  reconstruct the HDE model with GCG in the framework of KK cosmology. The letter is organized as follows: in section 2. field equations for KK cosmology are introduced. Then in section 3. we propose the Interactive Holographic GCG model in KK cosmology. The stability of the model is discussed in section 4. and finally in section 5 a brief discussion is presented. 
\section{Field Equations of Kaluza-Klein Cosmology}
\label{feqkk}
The Einstein field equation is:
\begin{equation}
\label{efeq}
R_{AB}-\frac{1}{2}g_{AB}R=\kappa T_{AB}
\end{equation}
where $A$ and $B$ runs from $0$ to $4$, $R_{AB}$ is the Ricci tensor, $R$ is the Ricci scalar and $T_{AB}$ is the energy-momentum tensor.
The $5$-dimensional spacetime metric of Kaluza-Klein (KK) cosmology is given by \cite{ozel} :
\begin{eqnarray}
\label{kkmetric}
ds^2 &=& dt^2-a^2(t)\left[\frac{dr^2}{1-kr^2}+r^2(d\theta^2 \right. \nonumber \\
&& \left. + sin^2\theta d\phi^2\right)+\left(1-kr^2)d\psi^2\right],
\end{eqnarray}
where $a(t)$ denotes the scale factor and $k=0,1(-1)$ represents the curvature parameter for flat and closed(open) universe. We consider a cosmological model where KK universe is filled with a perfect fluid.  The Einstein's field equation for the metric given by (\ref{kkmetric}) becomes:
\begin{equation}
\label{kkfeq1}
\rho=6\frac{\dot{a}^2}{a^2}+\frac{k}{a^2},
\end{equation}
\begin{equation}
\label{kkfeq2}
p=-3\frac{\ddot{a}}{a}-3\frac{\dot{a}^2}{a^2}-3\frac{k}{a^2}.
\end{equation}
For simplicity, we consider a flat universe, i.e. $k=0$. The above equations (eq. (\ref{kkfeq1}) and eq. (\ref{kkfeq2})) reduce to:
\begin{equation}
\label{kkfeq3}
\rho=6\frac{\dot{a}^2}{a^2},
\end{equation}
\begin{equation}
\label{kkfeq4}
p=-3\frac{\ddot{a}}{a}-3\frac{\dot{a}^2}{a^2}.
\end{equation}
The Hubble parameter is defined as $H=\frac{\dot{a}}{a}$. $T^{\mu\nu}_{;\nu}=0$ yields the continuity equation:
\begin{equation}
\label{kkcont1}
\dot{\rho}+4H(\rho+p)=0
\end{equation}
Using the equation of state $p=\omega\rho$, the equation of continuity becomes:
\begin{equation}
\label{kkcont2}
\dot{\rho}+4H\rho(1+\omega)=0
\end{equation}
We consider two kinds of fluids with total energy density $\rho=\rho_{\Lambda}+\rho_{m}$, $\rho_{\Lambda}$ corresponds to dark energy and $\rho_{m}$ is for matter inclusice of Cold Dark Matter (CDM) with an equation of state parameter $\omega_{m}=0$ hold separately. For non-interacting fluid, the  conservation equations for $p_{\Lambda},\rho_{\Lambda}$ and $p_{m},\rho_{m}$. However, in the case of  interacting dark energy models we consider the following equations admitted by the continuity equation : 
\begin{equation}
\label{kkcontm}
\dot{\rho_m}+4H\rho(1+\omega_m)=Q
\end{equation}
\begin{equation}
\label{kkcontl}
\dot{\rho_{\Lambda}}+4H\rho(1+\omega_{\Lambda})=-Q
\end{equation}
where $Q$ denotes the interaction between dark energy and dark matter.

\section{Interacting Holographic GCG model}
\label{int GCG}
Consider the interaction quantity $Q$ to be of the form $Q=\Gamma \rho_{\Lambda}$ and denote the ratio of the energy densities with $r$, i.e. $r=\frac{\rho_m} {\rho_{\Lambda}}$. This gives the decay of GCG into CDM and $\Gamma$ is the decay rate. Following ref. \cite{setarehol} we define effective equation of state parameters as follows:
\begin{equation}
\label{efeosdef}
\omega^{eff}_{\Lambda}=\omega_{\Lambda}+\frac{\Gamma}{4H} \; \; and \; \; \omega^{eff}_{m}=-\frac{1}{r} \frac{\Gamma}{4H}
\end{equation}
The continuity equations are given by:
\begin{equation}
\label{kkcontm2}
\dot{\rho_m}+4H\rho(1+\omega^{eff}_m)=0
\end{equation}
\begin{equation}
\label{kkcontl2}
\dot{\rho_{\Lambda}}+4H\rho(1+\omega^{eff}_{\Lambda})=0
\end{equation}
In terms of Hubble parameter the Friedmann equation in flat KK universe can be rewritten as:
\begin{equation}
\label{frwkkh}
H^2=\frac{1}{6} \left(\rho_{\Lambda}+\rho_m \right),
\end{equation}
where we consider $M_p^2=1$. Density parameters are defined as:
\begin{equation}
\label{dpdef}
\Omega_m=\frac{\rho_m}{\rho_{cr}},  \Omega_{\Lambda}=\frac{\rho_{\Lambda}}{\rho_{cr}},
\end{equation}
where $\rho_{cr}=6H^2$. In terms of density parameters equation (\ref{frwkkh}) becomes:
\begin{equation}
\label{frwdp}
\Omega_m+\Omega_{\Lambda}=1
\end{equation}
Using eqs. (\ref{dpdef}) and (\ref{frwdp}) we obtain:
\begin{equation}
\label{rdp}
r=\frac{1-\Omega_{\Lambda}}{\Omega_{\Lambda}}
\end{equation}
We consider here GCG for describing interacting dark energy model (EOS is given by eq. (\ref{gcgeos})). For GCG in KK cosmology the energy density is:
\begin{equation}
\label{rhoa}
\rho_{\Lambda}=\left[ A+\frac{B}{a^{4(1+\alpha)}} \right]^\frac{1}{1+\alpha}
\end{equation}
The EOS parameters  $\omega$ are  given by:
\begin{equation}
\label{oml}
\omega_{\Lambda}=\frac{p_{\Lambda}}{\rho_{\Lambda}}=-\frac{A}{\rho^{\alpha+1}_{\Lambda}}=-\frac{A}{A+\frac{B}{a^{4(1+\alpha)}}},
\end{equation}
\begin{equation}
\label{omlef}
\omega^{eff}_{\Lambda}=-\frac{A}{A+\frac{B}{a^{4(1+\alpha)}}}+\frac{\Gamma}{4H}.
\end{equation}
Now we consider a holographic correspondence for GCG in KK cosmology. For that as shown in \cite{sharifkk} the energy density for flat KK universe :
\begin{equation}
\label{rhohol}
\rho_{\Lambda}=3c^2\pi^2L^2.
\end{equation}
The infrared cutoff of the universe $L$ in the flat KK universe is equal to the apparent horizon, which coincides with Hubble horizon \cite{cai}. So we can write as in \cite{sharifkk}:
\begin{equation}
\label{cutoff}
r_a=\frac{1}{H}=r_H=L
\end{equation}
The decay rate \cite{wang} is now given by  ,:
\begin{equation}
\label{decay}
\Gamma=4b^2(1+r)H
\end{equation}
Using eq. (\ref{decay}) in eq. (\ref{omlef}) we obtain:
\begin{equation}
\label{omleffin}
\omega^{eff}_{\Lambda}=\frac{b^2-2}{1+\Omega_{\Lambda}}.
\end{equation}
The correspondence between holographic dark energy and GCG in KK model enables us to equate  eq. (\ref{rhoa}) and eq. (\ref{rhohol}), which gives:
\begin{equation}
\label{A1}
\left[A+\frac{B}{a^{4(1+\alpha)}}\right]^{\frac{1}{1+\alpha}}=3c^2\pi^2L^2.
\end{equation}
Using   eq. (\ref{rhohol}) and eq.  (\ref{omlef}), we  compute $A$, which is given by :
\begin{equation}
\label{afin}
A=\frac{b^2-2\Omega_{\Lambda}}{\Omega_{\Lambda}(1+\Omega_{\Lambda})}(3c^2\pi^2H^{-2})^{1+\alpha}
\end{equation}
Once again using eq. (\ref{afin}) in eq. (\ref{rhoa}),  and comparing it with eq. (\ref{rhohol}) we determine  $B$, which is given by:
\begin{equation}
\label{bfin}
B = \frac{\Omega_{\Lambda}^2-\Omega_{\Lambda}-b^2}{\Omega_{\Lambda}(1+\Omega_{\Lambda})}(3c^2\pi^2H^{-2}a^4)^{(1+\alpha)}.
\end{equation}
\section{Squared Speed Of Sound in Chaplygin Gas and Stability of the Model}

\begin{figure*}

\subfloat[Part 1][Variation of $\omega^{eff}_{\Lambda}$]{\includegraphics[width=0.4\textwidth]{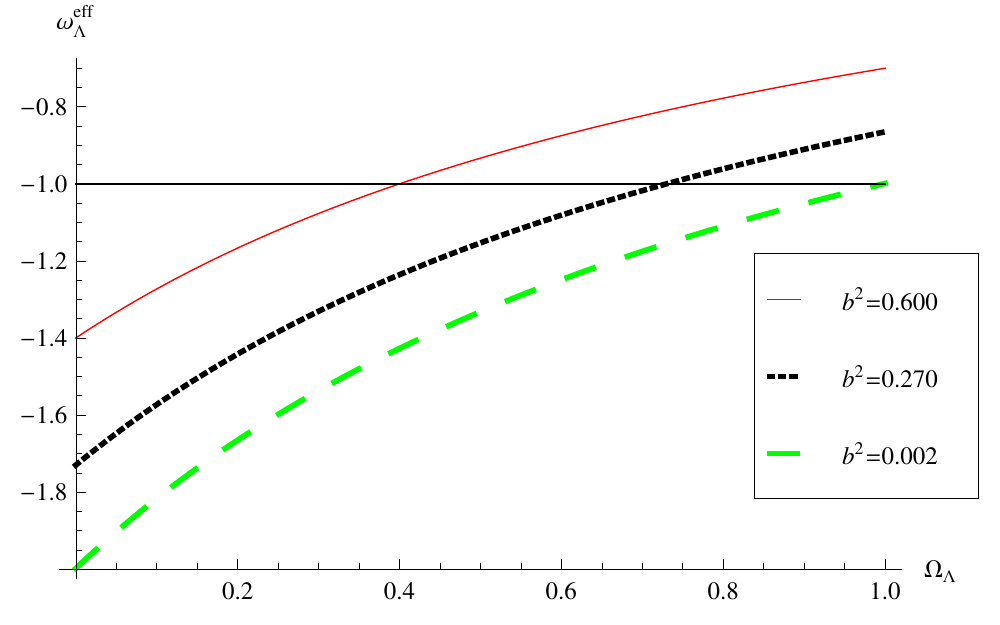} \label{oeff}} \qquad
\subfloat[Part 2][Variation of Sqared speed for GCG ($v_{\Lambda}$)]{\includegraphics[width=0.4\textwidth]{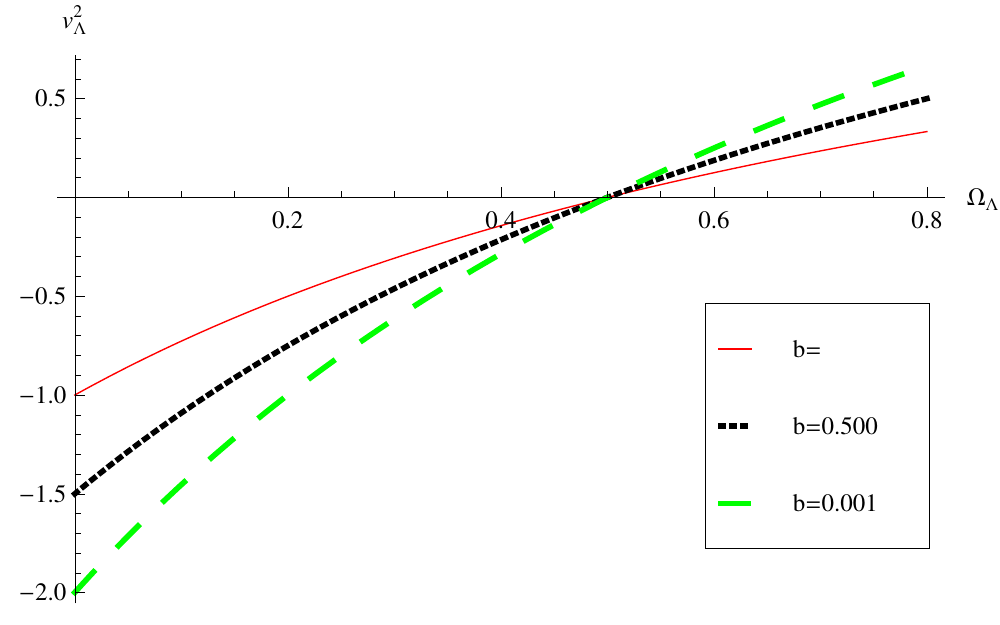} \label{sqvf}}\\
\caption[]{(Colour Online)Variation of (a) Effective dark energy equation of state and (b) squared speed for GCG is shown with $\Omega_{\Lambda}$For different choices of $b^2$ }
\end{figure*}

Stability of the GCG model may be analyzed determining the squared speed for GCG which is defined as:
\begin{equation}
\label{sqgcg}
v_g^2=\frac{dp_{\Lambda}}{d\rho_{\Lambda}}.
\end{equation}
The GCG model is unstable if $v_g^2 < 0$ \cite{setarehol}.  For our model of holographic interacting GCG:
\begin{equation}
\label{sqgcgchol}
v_{\Lambda}^2=\frac{dp_{\Lambda}}{d\rho_{\Lambda}}=\frac{\dot{p}}{\dot{\rho}}.
\end{equation}
In this case $\dot{p}$ is given by:
\begin{equation}
\label{dotp}
\dot{p}=\dot{\omega^{eff}_{\Lambda}}\rho_{\Lambda}+\omega^{eff}_{\Lambda}\dot{\rho_{\Lambda}}
\end{equation}
where the over dot means differentiation with respect to time. So the expression for squared speed becomes:
\begin{equation}
\label{sqv}
v_{\Lambda}^2=\omega^{eff}_{\Lambda}+\dot{\omega^{eff}_{\Lambda}} \frac{\rho_{\Lambda}}{\dot{\rho_{\Lambda}}}.
\end{equation}
Using eq. (\ref{omleffin}) one would obtain:
\begin{equation}
\label{dotom}
\dot{\omega^{eff}_{\Lambda}}=-\frac{b^2-2}{(1+\Omega_{\Lambda})^2} \dot{\Omega_{\Lambda}},
\end{equation}
where $\dot{\Omega_{\Lambda}}$ is determined from  eqs. (\ref{dpdef}) and  (\ref{rhohol}). Using the relation $L=\frac{1}{H}$, we get:
\begin{equation}
\label{dotom2}
\dot{\Omega_{\Lambda}}=-2c^2\pi^2\frac{\dot{H}}{H^5}.
\end{equation}
Using the values for $\dot{\omega^{eff}_{\Lambda}}$ and $\dot{\Omega^{eff}_{\Lambda}}$ in eq. (\ref{sqv}), one   obtains:
\begin{equation}
\label{sqvfin}
v_{\Lambda}^2=\omega^{eff}_{\Lambda}-\frac{c^2\pi^2(b^2-2)}{H^4(1+\Omega_{\Lambda})}=\omega^{eff}_{\Lambda}(1-2\Omega_{\Lambda}).
\end{equation}
Recent  predictions from various observations puts a value on  $\Omega_{\Lambda}\approx 0.73$, which will be used here. From eq. (\ref{sqvfin}) it would give $v_{\Lambda}^2=-0.46\omega^{eff}_{\Lambda}$. But $\omega^{eff} <0$ since it represents a form of dark energy, which ensures that $v_{\Lambda}^2>0$. Thus it is evident that our GCG model is stable  at the present epoch.

\section{Discussion}

In this letter,  a holographic dark energy (HDE) model is obtained considering interacting generalized Chaplygin Gas in KK Cosmology. Recently, Sharif and Khanum showed that interacting dark energy models can be proposed in the framework of compact KK cosmology. It is  also shown that generalized second law of thermodynamics (GSLT) holds good at all time in the HDE model considered here.  The above properties holds good in the case of interacting dark energy model described by GCG. It is seen from eq. (\ref{omleffin}) that the decay rate or the interaction term may be computed at the present epoch  for a given $\omega^{eff}_{\Lambda}$. For example if we chose $\omega^{eff}_{\Lambda}=-1$ at the present epoch it corresponds to $b^2=0.27$. Naturally for $b^2<0.27$ the dark energy behaves like  phantom. Since $\Omega_{\Lambda}$ is different for different epoch $b^2$ will take different values. Thus the dark energy might have evolved from a phantom phase at early epoch. This is evident from fig. (\ref{oeff}). It is to note that  HDE model for Interacting GCG is unstable in standard GR (as shown by Setare in \cite{setarehol} ) but in KK cosmology a stable model for the same is possible at the present epoch. This is evident from fig. (\ref{sqvf}) where stability is studied by a plot of $v_{\Lambda^2}$ (squared speed)  with $\Omega_{\Lambda}$ for various choices of $b^2$.The squared speed of sound in the case of GCG is not very much sensitive to $b^2$ values and becomes positive at around $\Omega_{\Lambda}\approx0.5$ for a range of $b^2$ values.

\begin{acknowledgement}
SG thanks CSIR for providing Senior Research Fellowship. BCP is thankful to IUCAA for providing visiting Associateship where the a part of the work is carried out. AS and RD are thankful to IRC, Physics Department, NBU for hospitality and extending facilities to complete the work.
\end{acknowledgement}

\bibliographystyle{spphys}
\bibliography{refs}

\end{document}